\documentclass[preprint,12pt]{elsarticle}

\usepackage{xcolor}
\usepackage{subcaption}

\usepackage{amssymb}%
\usepackage{hyperref}%

\usepackage{amsmath}%

\usepackage{tikz}

\usepackage{tabularx,siunitx,booktabs}%
\usepackage[linesnumbered,ruled]{algorithm2e}%

\usepackage{geometry}%
 \biboptions{sort&compress}%
 \usepackage{array,multirow,graphicx}

 \usepackage{lineno}

\begin{document}

\begin{frontmatter}		
% 	\title{Impact of Ecological Risk Factors on the Dynamics of Alcohol Drinking Amongst ASU Students: A simulation based model Implemented in NetLogo}
 	 	\title{The Impact of Individual's Ecological Factors on the Dynamics of Alcohol Drinking among Arizona State University Students:\\ {\em An Application of the Survey Data-driven Agent-based Model}}
 \author[add1]{Asma Azizi}
 \ead{aazizibo@uci.edu}
    \author[add2]{Anamika Mubayi}
     \author[add3]{Anuj Mubayi}

 \address[add1]{Department of Mathematics, University of California Irvine, Irvine, CA, USA}
 \address[add2]{ Department of Chemistry
University of Allahabad
Allahabad, UP-211002, India}
 \address[add3]{Simon A. Levin Mathematical, Computational, and Modeling Sciences Center, School of Human Evolution and Social Change, Arizona State University, Tempe, AZ, USA}
\begin{abstract}
%background:
 College-aged  students are one of the most vulnerable populations to high-risk alcohol drinking  behaviors that could cause them consequences such as injury or sexual assault. An important factor that may influence college students' decision on alcohol drinking behavior is socializing at certain contexts across university environment. The present study aims to identify and better understand ecological conditions driving the dynamics of distribution of alcohol use among college-aged  students.
  %method:
  To this end, a pilot study is conducted to evaluate students'  movement patterns to different contexts across the Arizona State University (ASU) campus, and  to use its results  to develop an agent-based simulation model designed for examining the role of environmental factors on development and maintenance of alcohol drinking behavior by a representative sample  of ASU students. The proposed  model that resembles an approximate reaction-diffusion model accounts for movement of agents to various contexts (i.e. diffusion) and alcohol drinking influences within those contexts (i.e., reaction) via a SIR-type model.
%result
%We use this model and simulations to conduct sensitivity analysis on context-related parameters.
Of the four most visited contexts at ASU Tempe campus -Library, Memorial Union,  Fitness Center, and Dorm- the context with the highest visiting probability, Memorial Union, is the most influential and most sensitive context (around $16$ times higher impact of alcohol related influences than the other contexts) on spreading alcohol drinking behavior. 
%Having the size of susceptible and at-risk pool increased by the number of agents per time in this context, the number of students exposed to drinking behavior would be higher without any intervention  for depleting this pool.
%conclusion:
 Our  findings highlight the crucial role of  socialization at  local environments on the  dynamics of students' alcohol use  as well as on the long-term prediction of the college drinking prevalence.
\end{abstract}

\end{frontmatter}

%\tableofcontents
\clearpage
\newpage
	\section{Introduction}
	Alcohol drinking among college students remains a key public health issue.
Approximately $62-72\%$ of college students in United States drink
at least once a month \cite{o2005predicting,kypri2005episode}, with an average number of drinking days per
month  ranged in $4-10$ \cite{fishburne2006college, burden2000expectancies, cameron2006stepping}, while over $40\%$ of which turns into heavy drinkers \cite{dantzer2006international}. According to Commission on Substance
Abuse at Colleges and Universities (CASA), increased drinking behavior among college students can raise violence and sexually transmitted infections \cite{malloy1994rethinking}, and according to the Global Status Report on Alcohol and Health it indirectly causes more than $60$ major types of diseases and injuries leading to  approximately $100,000$ deaths a year \cite{nutt2010drug}.
Besides, alcohol has a significant  effect on students academic performance and  social behaviour \cite{aertgeerts2002relation,singleton2009alcohol}.
The motives of college students for drinking  are often conformity motives (drinking to avoid social rejection), enhancement motives (drinking for enjoyment),  social motives (drinking to increase social rewards) or  coping motives \cite{cooper1994motivations, norberg2010social,kuntsche2005young,park2004positive}, and it can be associated with social anxiety \cite{buckner2006social, stewart2006relations,norberg2009refining}.

Quantitative complex systems modelling approaches can contribute to raise an insight  for policy making and evaluation \cite{purshouse2018Commentary}. Apostolopoulos et al. \cite{apostolopoulos2018moving1, apostolopoulos2018moving2}
 developed a case study of  quantitative
complex systems modelling  for   alcohol prevention
research  to display the  potential usefulness of such models  for policy making in the area of alcohol prevention.
 In this direction, a wide range of  epidemiological-type  mathematical models have been developed to  study the transmission dynamics and evolution of alcohol consumption  among college students and to evaluate various interventions, \cite{benedict2007modeling, manthey2008campus, huo2012global, mubayi2011types, mubayi2010impact,xiang2016modelling, scribner2009systems} and the references herein. These models assume that alcohol drinking is a contagious phenomenon spread by social interaction with alcoholic individuals or affected by social norms via aggregate parameters and influences in social circles. Therefore, these models  are  not able to look at individual decisions  creating these  aggregate phenomena \cite{epstein1996growing, garrison2009alcohol}. A more complex approach is needed to capture the dynamic complexity of college student alcohol use  and help college administrators or   community leaders make college drinking prevention strategies \cite{apostolopoulos2018moving1}.  Agent based computer models (ABM)- that are subject to random events- are useful to capture micro-level (each individual  own characteristic and decision) and macro-level (influence of environment and aggregated population) aspects of a typical social issue \cite{windrum2007empirical}. There have been numerous ABMs devoted to understand alcohol drinking behaviors and to provide intervention approaches to control it \cite{garrison2009alcohol, gorman2006agent,ip2012agent,fitzpatrick2016effectiveness, braun2006applications}. Garrison et al. \cite{garrison2009alcohol} provided a simulation tool to evaluate impact of peer pressure on alcohol drinking among students in an artificial society. Gorman et al. \cite{gorman2006agent} designed an ABM to examine the impact of environment interaction on drinking behavior at population level. They found  a tipping point  mixing rate beyond which the conversion rate of  nondrinkers was saturated. The existence of a leverage point for intervention targeting heavy drinkers or providing school policies \cite{ip2012agent} or making campaigns to revise the solution norms \cite{fitzpatrick2016effectiveness} was discussed via ABM as well. 

 The  environment influences individual’s decision on  alcohol consumption \cite{cox1988motivational}, and students in the various locations of university campus (called social context) may become exposed to social activities   associating  with an increase in risky alcohol use \cite{read2010before, schulenberg2002developmental}. On the other hand,  the chance of socialization at various contexts plays an important role on  individual's changing norms for  adapting to their social network, for example what they have previously  have thought was a heavy amount of drinking could become the new standard \cite{caldeira2012cigarette}.
In spite of importance of socializing context, none of the mentioned  models discusses its direct impact on alcohol drinking dynamics, the main focus of this study.

{\em Here, we use an spatial dynamical model to  understand the social environmental mechanisms that drive the spread of alcohol drinking behavior among students in the ASU campus. The understanding of dynamics may help prevent unhealthy behaviors of individuals later in their life leading to lower rates of preventable diseases and mortality \cite{witkiewitz2007modeling}.
We  use an  ABM that captures (a) students movement pattern in the campus area, (b) their interactions with others within different contexts and (c) each context's influences on students  alcohol consumption. The autonomous interacting agents of our model are students in an artificial campus area resembling Arizona State University, Tempe campus.}

Arizona Sate University (ASU) is one of the largest  public universities by enrollment in  United States with  an enrollment of more than $10,000$ students per year \cite{ASU}. Around $53\%$ of students at ASU consumed alcohol in the past $30$ days and $20\%$ of them drank heavily in their recent socializing gathering \cite{american2019american}. 
Because of the large and spread out ASU campus area in Tempe that provides access to several socializing contexts, we consider the contiguous campus building area as a case study of our work to specifically % address following questions
{\em 
\begin{enumerate}
\item design and collect cross sectional survey data to study correlation between  daily social movement in social contexts in the campus, environmental factors and current drinking patterns,
\item  
develop and use Classification and Regression Trees (CART) models to find clusters of student drinking population based on social and demographic factors,
\item develop an survey-based data-driven ABM framework for a college campus that captures  and links environmental (that is, contextual) influences and  demographic factors with the spatio-temporal dynamic of alcohol drinking  behaviors, and
\item estimate the mean drinking influence level for top four social contexts (in terms of number of visitors in a context) within the university and use it to identify the most vulnerable contexts to drinking initiation.
\end{enumerate}
}

\section{Method}\label{method}
    %Students in the campus may be encountered the situation-caused by friends (peer pressure) or environment -  where influence them to drink alcohol.
    This Section firstly describe and summarize the data collected from a students' survey carried out at ASU campus and secondly discusses development of the agent based model to study impact of university environment on students alcohol drinking behaviors. 
    
  %  collects  we  discuss on the methodology of how different environment agents  can affected individual agents.
    
    \subsection{Data sources}
    The IRB approved data was collected via online survey given to the college students at ASU, Tempe capmus. The population of this capmus is $52,000$, with  more than $50\%$ are younger than  $25$ years old. The $N=538$ students  participated in the survey, and answered  questions  about their demographic information as well as their daily activities.  The overall objective of the survey was to identify the relationships between ASU students  movement in the campus  and their alcohol drinking behavior. 
    
    The  students were asked about their (i) age,  (ii) gender, (iii) race (\# $1$-$9$; Asian, Hispanic/Latino, White/Caucasian, Black American, Black American/White, American Indian, South Asian, middle eastern, Mixed),  (iv) current year in school (\# $0$-$4$; Freshman, sophomore, junior,  senior, and graduate), (v)  participation in Greek life  (\# $0$-$3$ ; none; fraternity; sorority and others such as service, business, honors, etc; sorority and Co-ed business honors organization; professional, coed fraternity etc.), (vi) their daily activity (From the provided list of places, order them  from  the most visited to the least visited one at a typical work day), (vii) their alcohol drinking behaviors ( in the past $30$ days how many times they drank (\# $0-10$; No drink to $9$ or more times), and  (viii)  friends alcohol drinking behavior (How many close friends who are also  students of ASU they  have and how many of them  they consider as alcohol drinker). 
     
    \subsubsection{Data Exploratory Analysis} \label{DV}
\begin{figure*}[t!]
    \centering
    \begin{subfigure}[t]{0.5\textwidth}
        \centering
        \includegraphics[width=\textwidth]{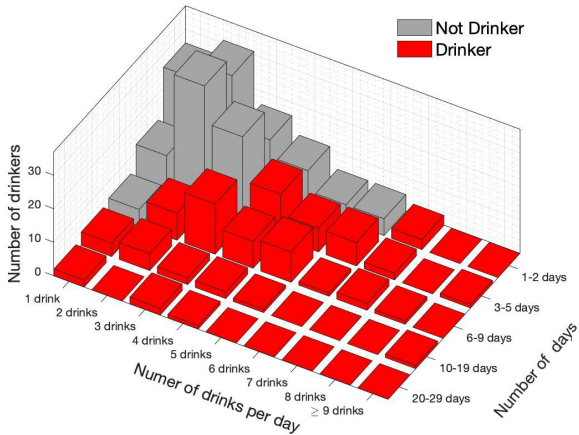}
        \caption{Drinking behavior of participants}
        \label{dp}
    \end{subfigure}%
\hfill
    \begin{subfigure}[t]{0.5\textwidth}
        \centering
        \includegraphics[width=\textwidth]{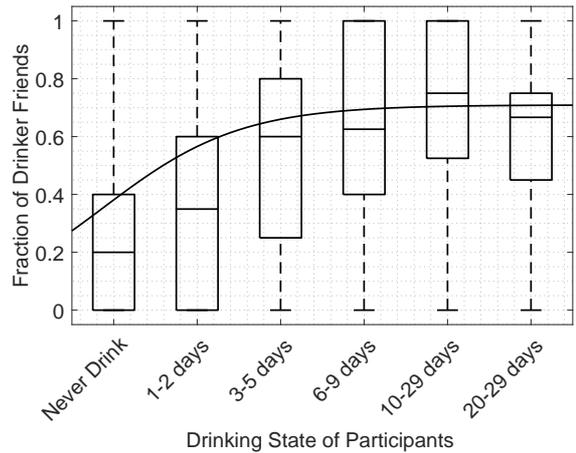}
        \caption{Drinking behavior of  participants and their friends}
        \label{dpf}
    \end{subfigure}
    \caption{\textbf{Drinking behavior of participants and their friends:} The left panel  is the two dimensional distribution of drinking state of participants who drink at least once in the last $30$ days, the red bars are  drinker participants (based on definition of drinker in  \cite{mubayi2008role}). Including participants who have never drank, the result shows around $20\%$ of participants are drinkers. The right panel is the correlation between drinking status of all participants and their friends, the y-axis it the fraction of drinker friends (based on the participants perception of drinker) and the solid curve is the fitted curve to median revealing of the existence of a tipping point, that is, for non or light drinkers, people who drink more have more drinker friends, but for heavy drinkers the fraction of drinker friends saturate to a fixed value of $70\%$.}
    \label{fig:drinking_data}
\end{figure*}

    The Table (\ref{tab:da}) summarizes  the demographic profile of the participants and the Figure (\ref{fig:drinking_data}) visualizes the participants drinking behavior and its correlation with their friends drinking behavior. The Figure (\ref{dp}) is the 3D bar plot of number of days (settings) and number of drinks per day (setting) for the participants who drank alcohol during the last $30$ days. 
    If we define a person as drinker if he/she drinks $3-4$ days (setting) per week or consumes $5$  or more drinks at any one day (sitting) per week \cite{mubayi2008role}, then around $20\%$ of the participants are categorized as drinkers, the red bars in Figure (\ref{dp}) noticing that participants who never drank were not showed in this plot.
    The Figure (\ref{dpf}) is the box plot of correlation between drinking behavior of all participants and the fraction of their drinker friends.  We observe a logistic shape for the median of the friction of drinkers friends, therefore, we fitted a logistic curve to it, the solid curve. This  fit  suggests that there is a threshold  effect  at $3-5$ days per month, that is,  for non or  light drinker participants (ones who  drink less than $3-5$ days per month) the ones  who drink more are more surrounded by drinker friends. After $3-5$ days per month, this trend will saturate to a fix value, that is, when the number of drink days goes beyond  this threshold the median for  the fraction of drinker friend saturate to fix value of around $70\%$.  This observation suggest that students who are not drinkers are more prone to become drinkers if they socialize with more drinker friends.
\begin{table}[htp]
\centering
\addtolength{\tabcolsep}{3pt} 
\resizebox{\columnwidth}{!}{\begin{tabular}{p{4cm}p{4cm}p{4cm}p{4cm}}
\toprule
\textbf{Variable} &{\textbf{Frequency }}&\textbf{Variable} &{\textbf{Frequency} }\\
&\textbf{(Percentage \pmb{\%})}&&\textbf{(Percentage \pmb{\%})}\\
\midrule
\textbf{Age} &&\textbf{Gender}&       \\
\hspace{.5cm}17 or less Years  &9 (1.7\%)&\hspace{.5cm}Male & 106  (19.7\%) \\
\hspace{.5cm}18-20 Years     &379 (70.4\%)&\hspace{.5cm}Female &428  (79.4\%) \\
\hspace{.5cm}21-25 Years    & 119 (22.1\%)&\hspace{.5cm}Non-Binary & 5    (0.9\%)
\\
\hspace{.5cm}25 or more Years  & 31 (5.8\%)&\textbf{Race }  & \\
\textbf{School Year }&&  \hspace{.5cm}White  & 287 (53.2\%)\\
\hspace{.5cm}Freshman    &163 (30.3\%) &  \hspace{.5cm}African-American   & 21 (3.9\%)\\
\hspace{.5cm}Sophomore  & 136 (25.3\%)  &\hspace{.5cm}Asian   & 108 (20\%)\\
\hspace{.5cm}Junior      & 125 (23.2\%) & \hspace{.5cm}Hispanic/Latino   & 73 (13.5\%)\\
\hspace{.5cm}Senior      &92 (17.7\%) & \hspace{.5cm}Other  & 9 (9.4\%)\\
\hspace{.5cm}Graduate    & 22 (4.1\%) & \\
\addlinespace

\midrule[\heavyrulewidth]
\end{tabular}}
\caption{\textbf{Demographic  profiles of participants}}
\label{tab:da}
\end{table}

\begin{figure}[htp]
    \centering
   \includegraphics[width=.8\textwidth]{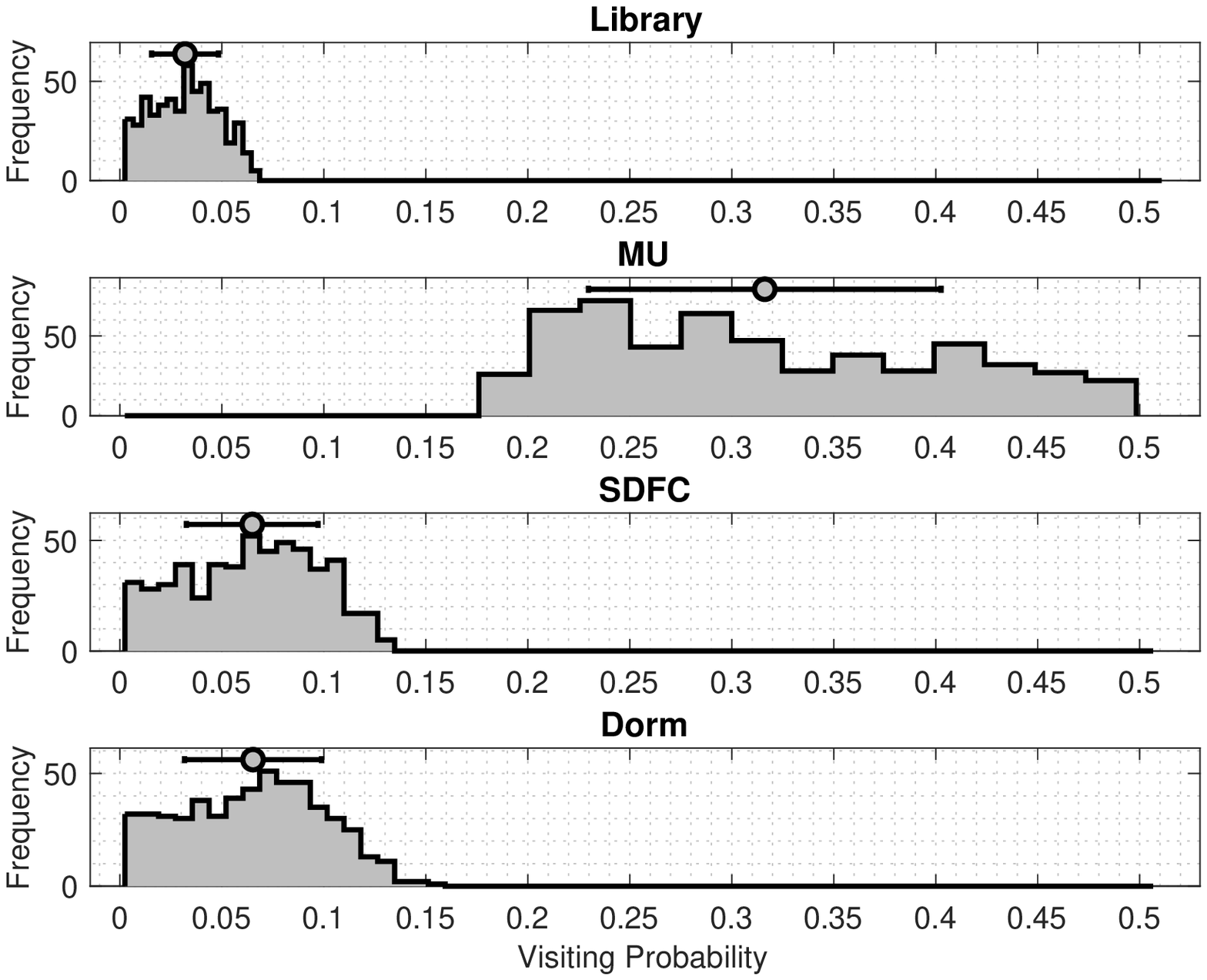}
    \caption{\textbf{Per day visiting probability of visiting  contexts}:  Each plot is the histogram for the number of participants (y-axis) who visit the specific context with  per unit time probability (x-axis). The circle on the top of plots is the mean of the data and the line passing mean shows the first and third quartiles of the data.  For instance $50$ participants visit MU with probability of $0.3$ and average probability of visiting MU for all participants is around $0.33$. MU is the most crowded location as its probability of visiting is higher than the other three contexts}. 
    \label{fig:visiting}
\end{figure}

Beside that demographic questions, participants were asked about the contexts they visit more often during the day. 
From this collected data, we found the four most visited contexts on the ASU campus as  Library (including Hyden and Nobel), Memorial Union (MU), Sun Devil Fitness Center (SDFC),  and dorms/off-campus housing (including three locations). Along with this, we have an extra context (called Others) that is every place but these four contexts.  The Figure (\ref{fig:visiting}) shows bar plot of visiting places based on the frequency. Among these four selected contexts, MU is the most visited one as the probability of visiting this context for all of the participants is higher than the other contexts.

\subsubsection{CART Analysis} 
Classification and Regression Trees (CART) analysis is a tree-based method used to make decision of how independent variables (predictor) explain dependent variables (target). The CART  technique   selects the most effective predictors and their interaction from a large number of variables. Here we addressed primary questions related to understanding of the raw data. 

We used CART on  our  data  to examine the association between alcohol drinking behavior measured by Setting, Drink per setting, and Drinking status (being a drinker or not) for all participant, the ones who live on campus area, and the ones who live off campus area.  Our  predictor variables are age, race, gender, the number of clubs belonged to, school year, and fraction of drinker friends. We summarize the all target and predictors  variables in Table (\ref{tab:cart}).
\begin{table}[htp]
 \centering
\resizebox{.85\columnwidth}{!}{
\begin{tabular}{llp{2cm}p{13cm}}
 \toprule[1.5pt]
    & \textbf{Symbol} & {\textbf{ Type} }&\textbf{Description}\\
  \cmidrule(lr){2-4}\cmidrule(l){2-4}
%     %=====================Target===============%
& Setting& Qualitative& Number of days used alcohol in the last month: 1-2 days, 3-5 days, 6-9 days, 10-19 days, 20-29 days\\
Target  &Drink.per.Setting& Integer& Number of drinks  per day $\in\{0,...,8,9^+\}$\\
Variables&Drinking Status&  Qualitative& Drinker, Not Drinker\\ 
 
 %=====================predictor===============%
 \cmidrule(lr){2-4}\cmidrule(l){2-4}
     &Age& Integer & Age of the participant\\  
 & Gender & Qualitative & F (Female), M (Male)\\
Predictor  & Race& Qualitative &  A(Asian), AA (African American), W(White),  HL (Hispanic/Latino),\\
Variables &&& AIAN (American Indian or Alaska Native), M (Mixed), O (Other)  \\
  & School.Year& Qualitative &  F(Freshman), SO (Sophomore), J(Junior),  SN (Senior), G (Graduate)  \\
  & Club & Integer & The number of clubs participant belongs to $\in \{0,...,5\}$\\
  & Drinker.Friends & Continuous & The fraction of drinker friends $\in [0,1]$\\
\bottomrule[1.5pt]
 \end{tabular}}
\caption{\textbf{The definition of variables used for CART analysis}}
 \label{tab:cart}
\end{table}
\begin{figure}[htp]
  \centering

  \begin{subfigure}{.9\linewidth}
    \centering
    \includegraphics[width = \linewidth]{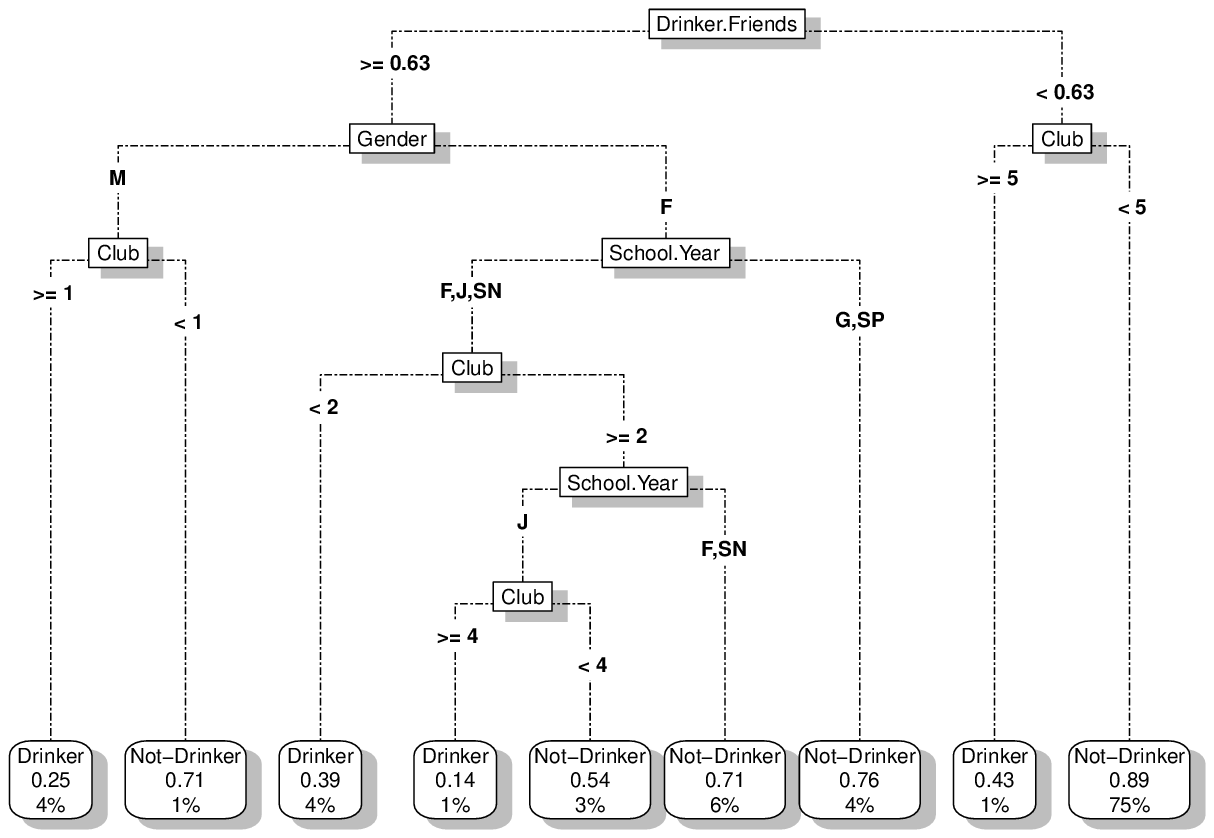}
    \caption{Drinking Status for all $538$ participants}
    \label{all}
  \end{subfigure}
    \begin{subfigure}{.45\linewidth}
    \centering
    \includegraphics[width = \linewidth]{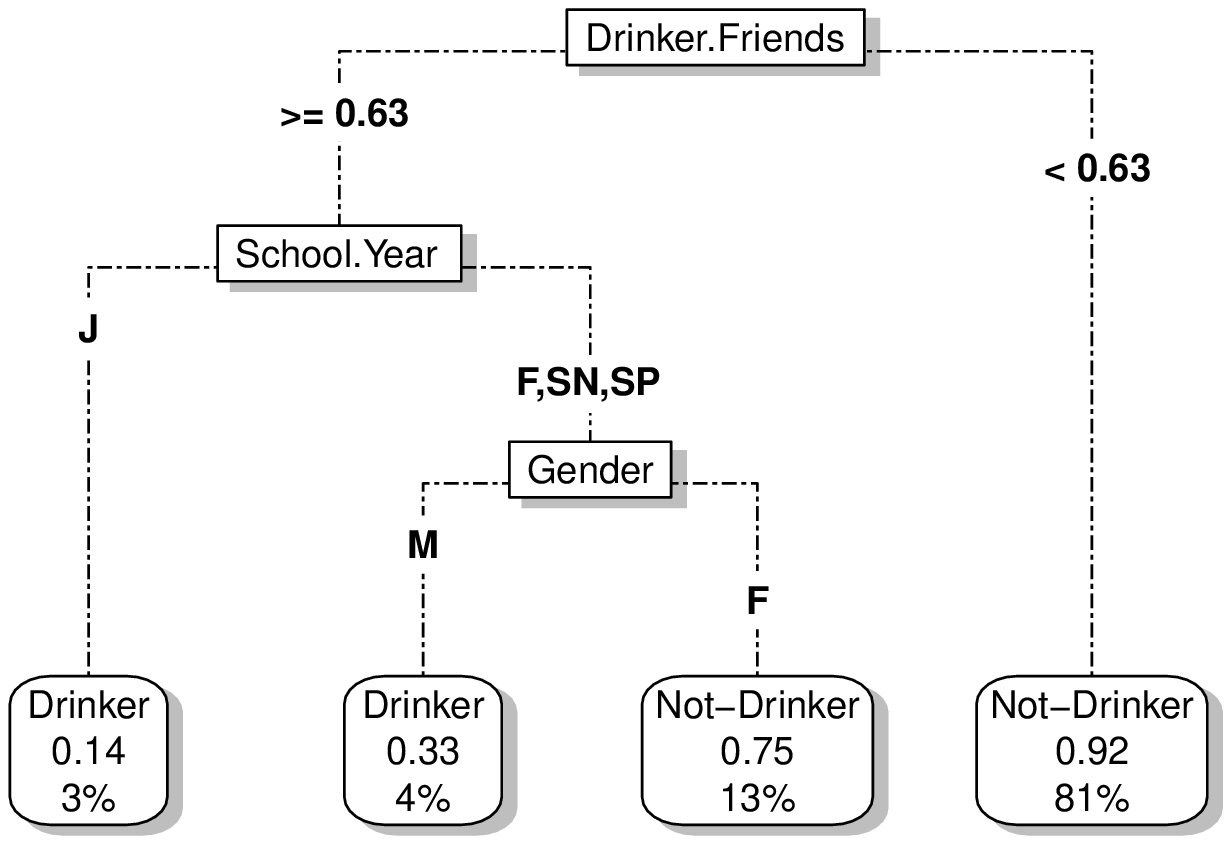}
    \caption{Drinking Status for  $269$ participants who live on campus}
    \label{on}
  \end{subfigure}%
   \begin{subfigure}{.45\linewidth}
    \centering
    \includegraphics[width = \linewidth]{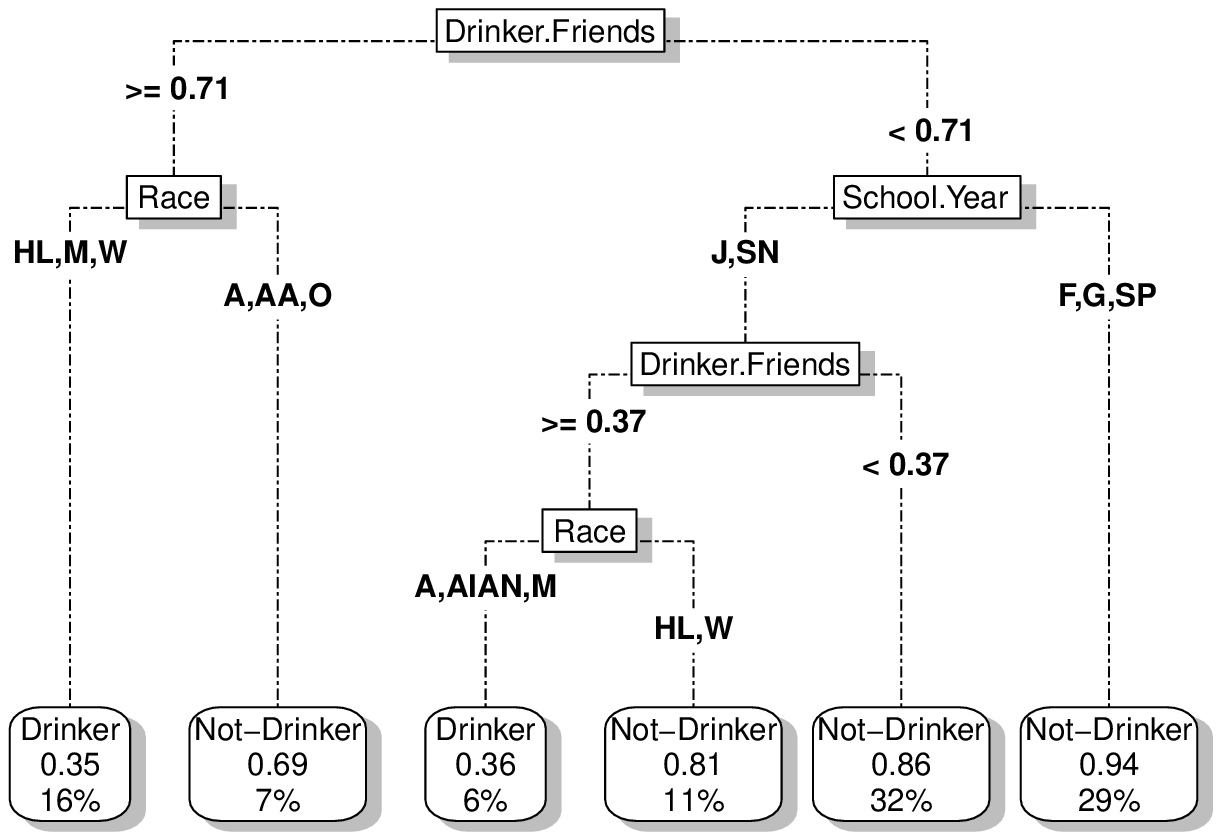}
    \caption{Drinking Status for $267$ participants who live off campus}
    \label{off}
  \end{subfigure}%
  \caption{\textbf{The output of CART analysis}: This plot includes the decision tree for drinking status for all participants (\ref{all}), participants who live on campus (\ref{on}), and participants who live off campus (\ref{off}). The last row of each tree represent the terminal nodes (clusters). Each terminal node includes the predicted drinking status (drinker or not-drinker), the probability of being not-drinker, and  the percentage of root sample size belonging to the corresponding category. For example based on the left terminal node in Subfigure (\ref{all}), $4\%$ of all participants are male who attend at least one club and more than $63\%$ of their friends are drinker. The probability of being not-drinker for these people is $0.25$, therefore, they are predicted as drinker.} 
  \label{fig:cart}
\end{figure}

 \begin{table}[htp]
 \centering
\resizebox{.83\columnwidth}{!}{
\begin{tabular}{llp{12cm}}
 \toprule[1.5pt]
  \textbf{Target }  & \textbf{Cluster} & {\textbf{ Description} }\\
  \textbf{Variable}  & \textbf{ Number} & \\
  \cmidrule(lr){1-3}\cmidrule(l){1-3}
%     %=====================setting===============%
&$3$&  Sixty one percent of the participants have less than $45\%$ of their friends as  drinkers, but themselves never drink.\\ 
\cmidrule(l){2-3}
   &$4$ &  Thirty percent of participants have  more than $45\%$  drinker friends but do not belong to $10\%$ of US race such as Asian and Pacific Islander and Native Hawaiian. They   drunk for on
an average of 3-5 days during past month. \\
\cmidrule(l){2-3}
Setting& $11$&Four percent of participants  have  more than $45\%$  drinker friends, are in 
minority race, and are in first, second or last year of their program. They, on average,  never  drunk  during past month.\\ \cmidrule(l){2-3}
 & $20$&  Three percent of participants  have  more than $45\%$  drinker friends, are in
minority race, and are and third year of their program. They, on average,    drunk  1-2 days during past month.  \\ \cmidrule(l){2-3}
 & $21$&  Three percent of participants  have  more than $45\%$  drinker friends, are in
minority race, and are in fourth year of their program. They, on average,    drunk  3-5 days during past month.  \\
   %=====================drink per setting===============%
\cmidrule(lr){1-3}\cmidrule(l){1-3}
&$2$&  Two percent of the participants are younger than $18$ years old and on average drunk $1.2$ times per setting during past month.  \\ 
\cmidrule(l){2-3}
   &$6$ &  Thirty percent of participants are older  than $18$ years old and belong to Asian, African American, Mixed or Other races. They  drunk for $3.4$ times per setting during past month. \\
\cmidrule(l){2-3}
& $14$& Twelve percent of participants are older  than $18$ years old and belong to minority or White races and attend at most one club. They  drunk for $3.3$ times per setting during past month.\\ \cmidrule(l){2-3}
 & $30$&  Thirty eight percent of participants are older  than $18$ years old and belong to minority or White races and attend more than one club and more than $10\%$ of their friends are drinker. They  drunk for $3.9$ times per setting during past month.  \\ \cmidrule(l){2-3}
 Drink.Per.Setting & $62$&  Six percent of participants are between $18$ and $19$ years old and belong to minority or White races and attend more than one club and less than $10\%$ of their friends are drinker. They  drunk for $3.7$ times per setting during past month.  \\
\cmidrule(l){2-3}
 & $126$&  Six percent of participants are older than $20$ years old and belong to minority or White races and attend more than one club and less than $10\%$ of their friends are drinker. They  drunk for $4.1$ times per setting during past month.  \\
\cmidrule(l){2-3}
& $254$& Five percent of participants are between $19$ and $20$ years old and belong to minority or White races and attend more than one and less than three club and less than $10\%$ of their friends are drinker. They  drunk for $5.7$ times per setting during past month.  \\ \cmidrule(l){2-3}
& $255$& Two percent of participants are between $19$ and $20$ years old and belong to minority or White races and attend more than three clubs and less than $10\%$ of their friends are drinker. They  drunk for $9.4$ times per setting during past month.  \\
\bottomrule[1.5pt]
 \end{tabular}}
\caption{\textbf{A summary of result of CART analysis for various target variables}}
\label{tab:cart}
\end{table}

\noindent The Figures (\ref{fig:cart})  show the  trees reproduced by CART for the target variable Drinking Status.  We interpret one of them in details, Subfigure (\ref{on}) and then interpretation of the rest would be similar. For the other target variables, we summarize the result in Table (\ref{tab:cart}). 

The Subfigure (\ref{on}) shows the tree reproduced by  CART  for Drinking status of participants living on campus as a qualitative variable. 
The tree has four terminal nodes with Drinker.Friend and School.Year are the primary splitters in the  tree. That is,  these predictors are critical in classifying drinking status. 
On the top of the tree (node 1), we have Drinker.Friends, therefore, the single best predictor to classify Setting is fraction of drinker friends.
CART further directs the participants having more than $63\%$ of their friends as drinker   to the left forming decision node  2, 
and directs the rest to the right forming terminal node 3. As indicated by this terminal node, if  less than $63\%$ of friends are drinker, the tree predicts that the participant  most likely never drinks ( $92\%$ out of $81\%$ of all data in this split).
CART further splits node 2 based on  participant's School.Year and directs it with  $J$  to the left, forming terminal node 4; directs the rest  to the right, forming decision node 5. Terminal node 4 predicts that on average  involved participants are drinker. CART continues to split node 5 based on gender into decision node 10 and terminal nodes 11.
Terminal nodes 10 and  11 indicate that conditioned on having more than $63\%$ of friends as drinker, being Freshman, Senior or sophomore  and  being female, the participant  most probably  are not drinker( $75\%$ out of $13\%$ of all data in this split), but being male,  he most probably is drinker ( $100\%-33\%=67\%$ out of $4\%$ of all data in this split).
%\textcolor{red}{any idea which one support more drinking: living on or off campus?}

 For all of these trees the fraction of drinker friends plays an important role on drinking behavior of participants that this  finding is consistent with the analysis results shown in Subfigure (\ref{dpf}), and many other studies such as \cite{hussong2003social, martin1993alcohol}. Other relatively significant factors driving alcohol behavior are Race, School.Year and Club.
 There is no control on Race or School.Year, nevertheless, Drinker. Friends and Clubs can be representative of level of socializing and its indirect impact (supported by our CART analysis) on alcohol drinking behavior\cite{borsari2001peer}. Various contexts within university campus provide various amount of opportunity to socialize and therefore various risk of becoming an alcohol drinker. The topic of our next Subsection is  to measure this indirect impact of socializing context via  developing  an ABM.
 
\subsection{Agent Based Model}
ABMs- computer based models- simulate the actions and interactions of
autonomous agents   representing the individuals of the modeled population. The short- and long- term activity of  the agents
can be compiled to obtain population-level measures of a biological system at a given scale \cite{yong2015agent}. 
Here we create an agent-based stochastic model (ABM) in which students are  classified  based on their drinking status and college social context. To explain our model we follow the structure in \cite{grimm2006standard}  called  ODD (Overview- Design concepts-Details)  protocol. %The model is then used to  understand spread of drinking behavior  among students in the presence of peer and contextual influences in ASU campus area in Tempe. 
\subsubsection{Overview}
 \noindent\underline{\textit{Purpose}}: The purpose of the model is to understand how spending time at different locations of university campus (called contexts)  affects the dynamics of drinking behaviour   among students.\\~\\
  \noindent\underline{\textit{State variables and scales}}:
The model comprises three hierarchical levels:  environment (campus geographical area), context (social activities/events), and individuals (students). 

\begin{itemize}
    \item 
The {\bf environment}  resembles the campus area, which is  around $660$ acres $4 \times 4$ grid  from Rail road-Rural road in South East to University Dr--Fifth avenue in North West, Figure (\ref{fig:context}). 
\begin{figure}[htp]
    \centering
    \includegraphics[width=.85\textwidth]{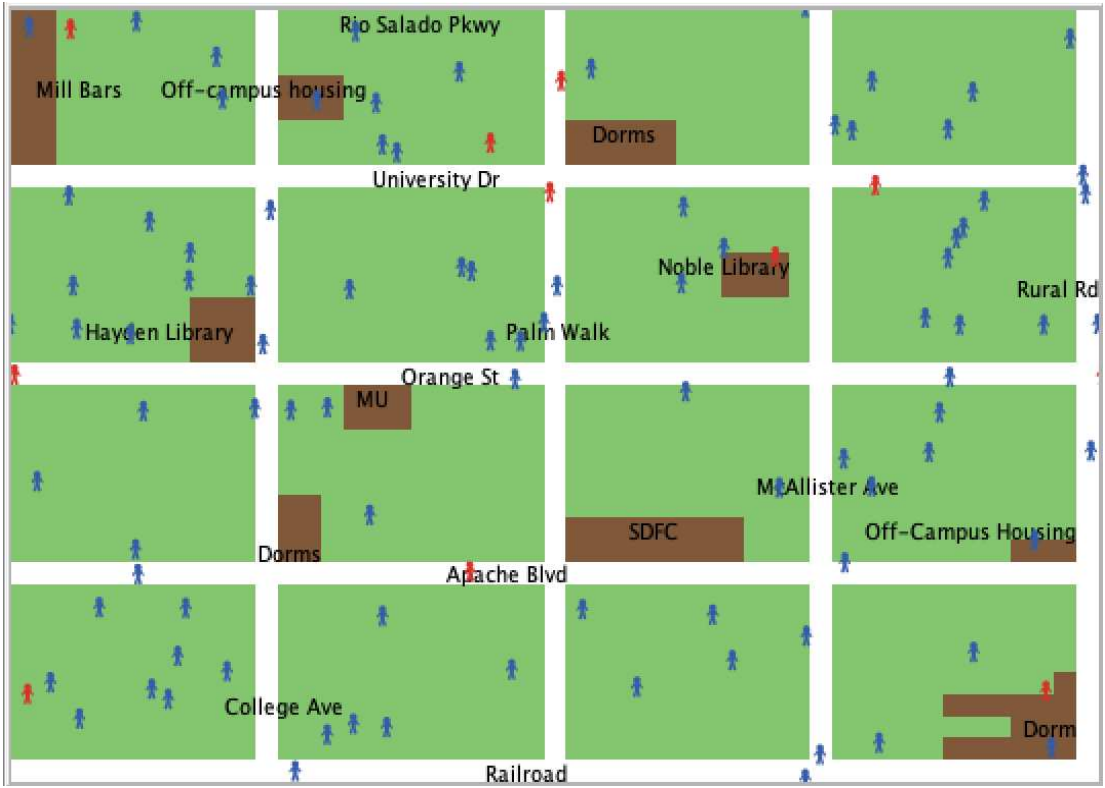}
  \caption{\textbf{The map  of ASU campus and spatial instance of simulation:} Four main contexts are the brown patches, and the rest is other area.}
    \label{fig:context}
\end{figure}
This environment includes several contexts -an specific location of university campus - in where individuals move around and interact. 

\item The {\bf contexts} are listed as
\begin{enumerate}
  \item Library: that includes two different libraries, Hyden that is the largest and most visited library facility at ASU, and Nobel library.
    \item Memorial Union (MU): considered as living room of campus. This context provides students with different experience activities such as  student organization,  community service, planning an event or eating  meal.
    \item Sun Devil Fitness Complex (SDFC):  also known as Physical Education West and is used mostly for intramural sports at the campus.
    \item Dorms: that are parts of Campus Housing. There are several residence halls (dorms) in Tempe campus.
    \item Other places: We assume if individuals are not spending time in the previous mentioned contexts, they are in other places shown in Figure (\ref{fig:context}) as long as they are in campus. 
\end{enumerate}
These contexts  are selected and labeled as $C_1-C_5$ as they are the most crowded contexts that individuals gather at events or collective activities, therefore, they are closer to each other rather than when they just move at random.  Each context  named $C_i$ for $i\in\{1,..,5\}$ is characterized by three  attributes. 
The first attribute is the constant probability of contact between two typical individuals visited context $C_i$, $\sigma_{C_i}$. To find this parameter, we used a  very ad hoc approach  assuming this probability is density-based with a power proportion behavior, that is, we define $\bar{N}_{C_i}$ as  per area  average number  of individuals visiting context $C_i$, then we have 
\begin{equation}\label{contact}
    \sigma_{C_i}=\frac{\sqrt{\bar{N}_{C_i}}}{\sum_{j=1}^4\sqrt{\bar{N}_{C_j}}},~~i=1,\cdots,4,
\end{equation}
and for the  last context $C_5$, we assume $\sigma_{C_5}=0$, that is, for the places other than the first four no contact causing to spread of drinking behavior happens.
 The function (\ref{contact}) is selected to represent  an intermediate between frequency- and density-dependence contacts \cite{mccallum2001should}, with  the square root that provides a saturating shape. Assuming the area of all four contexts under study is the same, we use the normalized mean probability of visiting context $C_i$ called $p_{C_i}$ (the circle points in Figure (\ref{fig:visiting})) and multinominal distribution to find $\bar{N}_{C_i}=Np_{C_i}$ as the  average number of individuals located in context $C_i$.
  The second attribute is constant probability of drinking behavior transmission within one contact happened at $C_i$. We call this probability as  drinking influence success $\beta_{C_i}$, which provides the chance that a drinker spread his/her behavior to others via a contact. The last attribute is an state variable of   the number  drinkers visited  context $C_i$ at time $t$, $I_{C_i}(t)$. These attributes for context $C_i$ are shown by $\theta_{C_i}(t)=(\sigma_{C_i},\beta_{C_i}, I_{C_i}(t))$

\item We consider $N=538$ (survey sample size) ASU college students in our ABM. An {\bf individual} (i.e., a typical student) in this cohort is represented by $\textbf{A}_i$,  $i\in\{1,..,N\}$, who is characterized by four attributes (some of them may change over time): 
\begin{itemize}
\item[(a)] Class-year, $R_i$: $0$ (Freshman), $1$ (Sophomore), $2$ (Junior),  $3$ (Senior), and $4$ (Graduate), 
\item[(b)] Probability of visiting a contexts, $P_i$, (the $i^{th}$ row of the following matrix):
$$\mathcal{P}_{N,n} = 
 \begin{pmatrix}
  P_1(C_1) & P_1(C_2)  & \cdots & P_1(C_n)  \\
  \vdots  & \vdots  & \vdots  & \vdots  \\
 P_i(C_1) & P_1(C_2)  & \cdots & P_i(C_n)  \\
  \vdots  & \vdots  & \vdots & \vdots  \\
  P_N(C_1) & P_N(C_2)  & \cdots & P_N(C_n)  
 \end{pmatrix}.$$
 The matrix $\mathcal{P}_{N,n} $ is designed and estimated based on our survey data. Hence, the matrix dimension is $538\times 5$ where % $\mathcal{P}(i,k)=P_i(C_k)$ 
 $P_i(C_k)$ entry is the probability of an individual $\textbf{A}_i$ to move to a context $C_k$, 
\item[(c)] the context that the individual $\textbf{A}_i$ is at time $t$, $V_i(t)$:  
(e.g., for our modeling population $V_i(t) \in \{ C_1, C_2, C_3, C_4, C_5 \}$,  $\forall t $), 
and 
\item[(d)]  drinking state $S_i(t)$, ($S_i(t) \in \{ ND, D, FD \}$,  $\forall t $)): The three  states are $ND$ (representing non-drinker individual who may be susceptible to initiating drinking behavior \cite{gorman2006agent}), $D$ (stands for current drinker, defined based on \cite{mubayi2008role} (see Subsection (\ref{DV}))), and $FD$ (stands for former drinker who were drinker before \cite{dawson2003methodological}).   
\end{itemize}

\sloppy To summarize, we define an attribute $\eta$  for individual  $\textbf{A}_i$ as  $\eta_i(t)=( R_i(t),P_i, V_i(t), S_i(t))$.
For example  $\eta_i(t)=(0, [0.14,0.06,0.41,0.24,0.15],C_k,ND)$ means that   non-drinker freshman individual $\textbf{A}_i$   has a $14\%$  chance of of being in context $C_1$,  $6\%$ chance in  context $C_2$ and so on, but currently  he/she is at context $C_k$.
\end{itemize}

\begin{figure}[hpt]
    \centering
    \includegraphics[width=.7\textwidth]{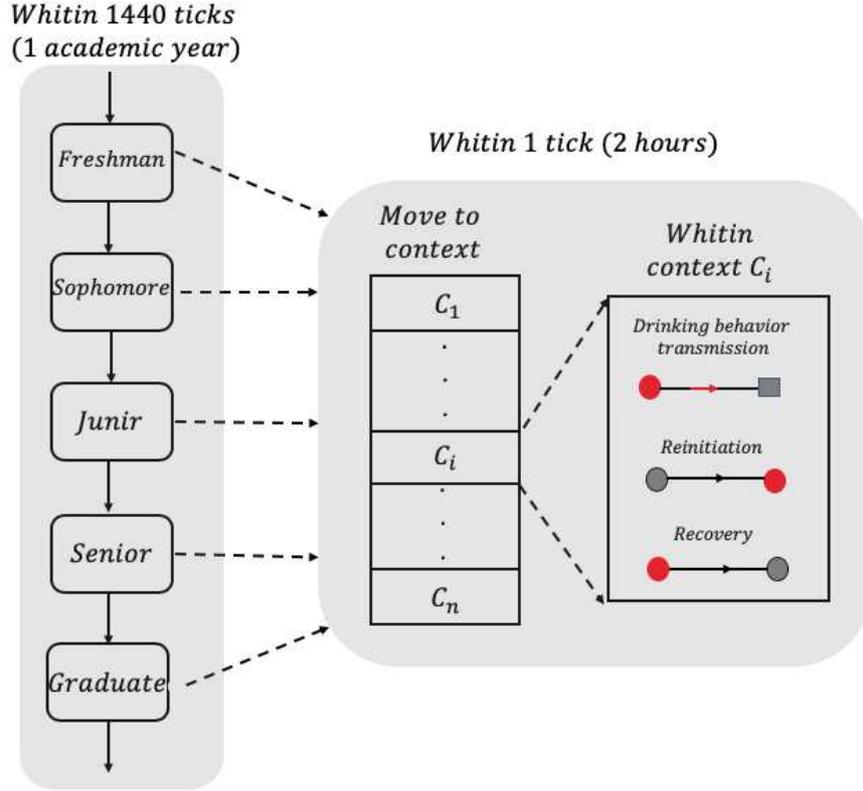}
  \caption{\textbf{Model phases}: Four phases are proceeding in every tick, and c phase is updated in every $1440$ tick.}
    \label{fig:abm_schedule}
\end{figure}

~\\~\\
\noindent\underline{\textit{Process overview and scheduling}}:
The model proceeds in two hours (1 tick) time steps with the assumption that a day is 16 hours (7am-11pm), or 8 ticks. Within each tick proceeding   phases are ordered as follow: individuals movement to various contexts, drinking behavior transmission, recovery from drinking, drinking reinitiation at individuals' discretion,  and class-year updates at the end of each $1440$ ticks (corresponding to one academic year or 9 months). At the end of academic year graduate  students  leave the school, but for each one graduate individual who leaves the school, we introduce a new freshman who is non-drinker, but all other their characteristic is the same as left one, therefore, the total number of individuals is conserved during the simulation.  Within each phase, individuals are processed in a random order. The phases 
are  depicted in Figure (\ref{fig:abm_schedule}).

\subsubsection{Design concepts}
\noindent\underline{\textit{Emergence:}} Total population dynamics emerge from the processes of update class-year. Transmission dynamics, however, partially emerges from the processes of movement and contacts. The number of individuals leaving   drinking state or reinitiate drinking  does not depend on the time  but is drawn from a proper  distribution that was observed in field studies, see Section (\ref{details}).
\\~\\
\noindent\underline{\textit{Stochasticity:}}  All movements, interactions and behavioural parameters are interpreted as probabilities, some drawn  from empirical probability distributions captured from data, while the rest are taken from some predetermined distributions. To summarize, the  stochasticity of model is the result of (a) movement to typical context $C_i$, (b) making contact with non-drinker and or former drinker and transmit drinking behavior by drinker,
(c) reinitiang drinking by former drinker at their discretion, and (d) recovery from drinking  state by drinker.   The Table (\ref{tab:stocastic}) in Appendix list all the random variables in our ABM and their corresponding distribution. 

\noindent\underline{\textit{Observation:}} To test our model, we observe the spatial distribution of  individuals, and  to analyse the model, we record context-related stochastic parameters and   population-level variables, i.e. group size distribution, time series of population size with various alcohol-related states: non-drinker, 
drinker, and former drinker.
\subsubsection{Details}\label{details}

 \noindent\underline{\textit{Initialization}}: At initial time we  randomly select $30\%$ as freshman, $25\%$ as sophomore, $23\%$ as junior, $18\%$ as senior and $4\%$ as graduate.  We also randomly select  $5\%$ of population as drinker and assume the rest are non-drinker. At the beginning of every  $8$ ticks (every day), individuals  are placed  randomly in the environment. 
 Each model simulation was run up to reaching quasi-stationary state for $200$ replicates.
  \\~\\
 \noindent\underline{\textit{Input}}:
     For the base simulation, the input parameters and probabilities were set so that the model fits  reported data and filed study.
      The total number of individuals $N=538$, total number of contexts $n=5$, and  movement probability vector of each individual is taken from survey data. Some drinking spread parameters (reinitiate drinking behavior $\rho$, leaving drinking $\gamma$) were found in literature \cite{gorman2006agent,sobell2000natural,dawson2005recovery}.
       The context influence rates $\beta_{C_i}$ are parameterised to resemble to current prevalence of drinking as a quasi-stationary state, that is, we calibrated these parameters to the current $20\%$ fraction of drinkers. The Table (\ref{tab:par-def-value}) list the  input parameters, their definition and baseline values.
\begin{table}[htp]
\centering
\resizebox{\columnwidth}{!}{\begin{tabular}{lp{8.5cm}ll}
\tabularnewline
\hline 
\textbf{Notation} & \textbf{Definition}& \textbf{Baseline}& \textbf{Ref.}\tabularnewline
\hline
$\beta_{Library}$& Drinking influence success at Library &$0.0105$
    &Calibrated\tabularnewline
$\beta_{MU}$& Drinking influence success at MU &$0.0033$&Calibrated\tabularnewline
$\beta_{SDFC}$& Drinking influence success at SDFC &$0.0073$&Calibrated \tabularnewline
$\beta_{Dorm}$& Drinking influence success at Dorm &$0.0074$&Calibrated \tabularnewline
$\sigma_{Library}$& Contact probability at Library &$0.1429$
    & Survey data\tabularnewline
$\sigma_{MU}$&Contact probability at MU &$0.4492$
   & Survey data\tabularnewline
$\sigma_{SDFC}$& Contact probability at SDFC & $0.2043$
    &Survey data\tabularnewline
$\sigma_{Dorm}$& Contact probability at Dorm &$0.2037$&Survey data\tabularnewline
$\rho$ & Reinitiation drinking at discretion probability&$ 0.0187$&\cite{gorman2006agent,sobell2000natural,dawson2005recovery}\tabularnewline
$\gamma$ & Recovery from drinking probability&$ 0.0187$&\cite{gorman2006agent,sobell2000natural,dawson2005recovery}\tabularnewline
\hline 
\end{tabular}}
\caption{\textbf{Parameters definition and  baseline value:} The model parameters describing the drinking behavior dynamic that are estimated using  survey data, or calibration method explained in Subsection (\ref{cab}), and or  obtained from the literature.
}
\label{tab:par-def-value}
\end{table}

      ~\\~\\
 \noindent\underline{\textit{Submodels}}:

\noindent Movement:  Through time  individuals move in the five different contexts within the environment. Each tick and for a typical individual $\textbf{A}_i$ with predetermined probability vector $P_i$, we generate random number $u\in [0,1]$. Then if  
   \begin{equation*}\label{l-context}
   \sum_{j=1}^{m-1}P_{i}(j)<u\leq\sum_{j=1}^{m}P_{i}(j),
   \end{equation*}
   for some  $1\leq m\leq n$, then we place $\textbf{A}_i$ in  context $C_m$.

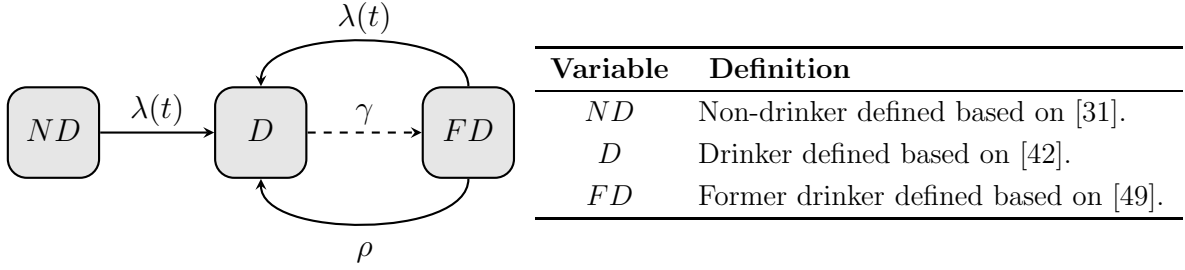
\begin{figure}[htp]
\begin{center}
\begin{minipage}[c]{0.35\textwidth}
\hspace{-2cm}
\tikzset{measurement/.style={circle, thick, minimum size=1.2cm, draw=orange!50, fill=orange!20},
input/.style={thick, ,rounded corners=0.25cm, minimum size=1.2cm, draw=black, fill=gray!20},
arrow/.style={thick,->,>=stealth}}
\begin{tikzpicture}[>=latex,text height=1.5ex,text depth=0.25ex]
\matrix[row sep=4em,column sep=0.75cm] (mat) {
&
 \node (ND) [input]{$ND$}; &
    &
    \node (D)   [input]{$D$};     &
    &
    \node (FD) [input]{$FD$}; \\   
 };
\begin{scope}[arrow]
 \foreach \X in {1} 
 {
 \draw (ND) -- (D) node[midway,above]{$\lambda(t)$};
  \draw [->,out=0,in=0,looseness=0, dashed ](D.east) to node[midway,above]{$\gamma$} (FD.west) ;
  \draw [->,out=90,in=90,looseness=0.75 ] (FD.north) to node[above]{$\lambda(t)$}  (D.north) ;
  \draw [->,out=-90,in=-90,looseness=0.75 ] (FD.south) to node[below]{$\rho$}  (D.south) ;
 }
\end{scope}
\end{tikzpicture}
\end{minipage}
\begin{minipage}[c]{0.5\textwidth}
\aboverulesep=0ex \belowrulesep=0ex \renewcommand{\arraystretch}{1.2}
\resizebox{\columnwidth}{!}{\begin{tabular}{cp{6.8cm}}
\toprule
\textbf{Variable }&\textbf{ Definition}\\
\midrule
 $ND$ &  Non-drinker defined based on \cite{gorman2006agent}. \\
$D$&Drinker  defined based on \cite{mubayi2008role}. \\
$FD$& Former drinker defined based on \cite{dawson2003methodological}. \\
\bottomrule
\end{tabular}}
\end{minipage}
\caption{\textbf{Role governing drinking behavior spread and variables definition}.}
\label{fig:model}
\end{center}
\end{figure}

 For  drinking behavior dynamic within any context we borrow the model in \cite{gorman2006agent} shown in   Figure (\ref{fig:model}).
\noindent This model  includes three processes explained below:
 
\noindent Drinking behavior transmission: On each tick, we model the ability for a drinker  individual
to spread  his/her behavior  to other non-drinker or former drinker visiting at the same context as follow: 
\begin{eqnarray*}
\resizebox{\textwidth}{!}{
$
\begin{aligned}
&\text{ Transmission}\\
&~\text{probability} 
\end{aligned}
= 
\left[
\begin{aligned}
&~~\text{Prob. of making a contact }\\
&~~~~~\text{ per individual}
\end{aligned}
\right]\times
\left[
\left(
\begin{aligned}
&\text{Prob. of drinking behavior }\\
&\text{transmission per tick per contact}
\end{aligned}\right)
\times
\left(
1- \text{Resistency}
\right)
\right]
 $}\\ [12pt]
% \resizebox{.2\textwidth}{!}{
\begin{aligned}
~~~=~~ ~~~~~~~ ~~~~~~~ &
\sigma_C~~~~~~~~~ ~~~~~
~~~\times~~~~~~~~~~~~~~~~~~~~
\beta_C~~~~~~~~~~~~~~~~
\times~~~~~~~~
\left(
1- \phi
\right),
%}
\end{aligned}
\end{eqnarray*}
where the resistancy $\phi$ is a measure for combination of  susceptibility level of non-drinker or former drinker individual  toward drinking behavior and infectivity level of drinker individual.   Because in real world, not all non-drinker or former drinkers are equally likely to become drinkers, and not all drinker are equally likely to transmit their behavior. Although the susceptibility of an individual may depends on its drinking state ( non-drinker or former drinker), for simplicity we assume non-drinker and former drinker both have the same susceptibility level.
With all these assumptions, we assume  resistancy $\phi$ follows a uniform distribution in  $[0,1]$, that is, for each contact between  a drinker and non-drinker (former drinker) individuals we assign a uniform random number where closer to zero means the contact is  less resistant and drinking behavior more likely to be transmitted and vice versa \cite{braun2006applications}. 

The aggregated probability that a  typical non-drinker  individual $\textbf{A}_i$  who visits context $C_k$ become drinker is 
\begin{eqnarray*}
P(S_{i}(t+\Delta t)=D&\vert& S_i(t)=ND~\& ~V_i(t)=C_k)=1-\left[1-\sigma_{C_k}\beta_{C_k}(1-\phi)\right]^{I_{C_k}(t)}.
\end{eqnarray*}
%\textcolor{blue}{
Therefore, the probability that non-drinker $A_i$ become drinker within time step $\Delta t$ is
\begin{equation*}
    \begin{aligned}
    \lambda(t)&=P(S_{i}(t+\Delta t)=D\vert S_i(t)=ND)=\sum_{k=1}^{5}P(S_{i}(t+\Delta t)=D \vert S_i(t)=ND~\& ~V_i(t)=C_k)\\
    &=\sum_{k=1}^{5} 1-\left[1-\sigma_{C_k}\beta_{C_k}(1-\phi)\right]^{I_{C_k}(t)}.
    \end{aligned}
\end{equation*}
%}
The former drinker has two reason  to become drinker: Transmission or Reinitiation at discretion.
The aggregated probability of becoming drinker  via transmission for a  typical former drinker $\textbf{A}_i$  who has not reinitiated drinking  at his/her discretion but visits context $C_k$ is
\begin{eqnarray*}
\begin{aligned}
P(S_{i}(t+\Delta t)=D&\vert& S_i(t)=FD~\&~ V_i(t)=C_k~\& ~\text{Transmission})
=1-\left[1-\sigma_{C_k}\beta_{C_k}(1-\phi)\right]^{I_{C_k}(t)}.
\end{aligned}
\end{eqnarray*}
\noindent Reinitiation at discretion:  The former drinker individual  $\textbf{A}_j$ who has not received drinking behavior by a drinker may  start drinking  independent of environment with probability
\begin{eqnarray*}
\begin{aligned}
P(S_{i}(t+\Delta t)=D&\vert& S_i(t)=FD~\&  ~\text{Re-initiation at discretion})
=\rho.
\end{aligned}
\end{eqnarray*}

%\textcolor{blue}{
That is, the total probability that former drinker $A_i$ become drinker within time step $\Delta t$ is
\[  P(S_{i}(t+\Delta t)=D\vert S_i(t)=FD)=  \left\{
 \begin{array}{ll}
      \lambda(t) & \textnormal{ Re-initiation due to Transmission} \\
      \rho & \textnormal{ Re-initiation at discretion.} \\
\end{array} 
\right. \]
%}

\noindent Leaving drinking state:  For a given drinker  individual $\textbf{A}_k$ the random variable $T_k> 0$ is defined to be the time at
which the $\textbf{A}_k$  changes its status to former drinker. This is modelled as being exponentially distributed so that the average time to recover is $\frac{1}{\gamma}.$ It then follows that $T_k$ is memoryless (probability that a drinker leaves drinking status to former drinker  is the same at each increment of time) and 
$$P(T
_k< \Delta t) = 1 - e^{-\gamma \Delta t}.$$

We summarized  all parameters of ABM in Table (\ref{tab:abm_par}) and   its full model process for a single tick $\Delta t$ in pseudocode (\ref{alg}) in Appendix.

 \section{Results}\label{result}
In this Section we carry out local and global sensitivity analysis to evaluate the
impact of context-related parameters involved in the model on drinking prevalence. 
Each simulation is  average of a bunch of single realizations of  stochastic ABM explained in Section \ref{method}. 
All of the realizations start at the same initial point obtained with the model baseline parameters in Table (\ref{tab:par-def-value}), unless stated otherwise.

\subsection{Model initialization}\label{cab}
We initialize our model based on estimates for the current drinking  epidemic among ASU students in Tempe campus.
The $5\%$ initial drinkers are  randomly distributed in an otherwise non-drinker population. They are distributed as they would be as part of an emerging epidemic that started sometime in the past.
\begin{figure}[htp]
    \centering
    \includegraphics[width=0.79\textwidth]{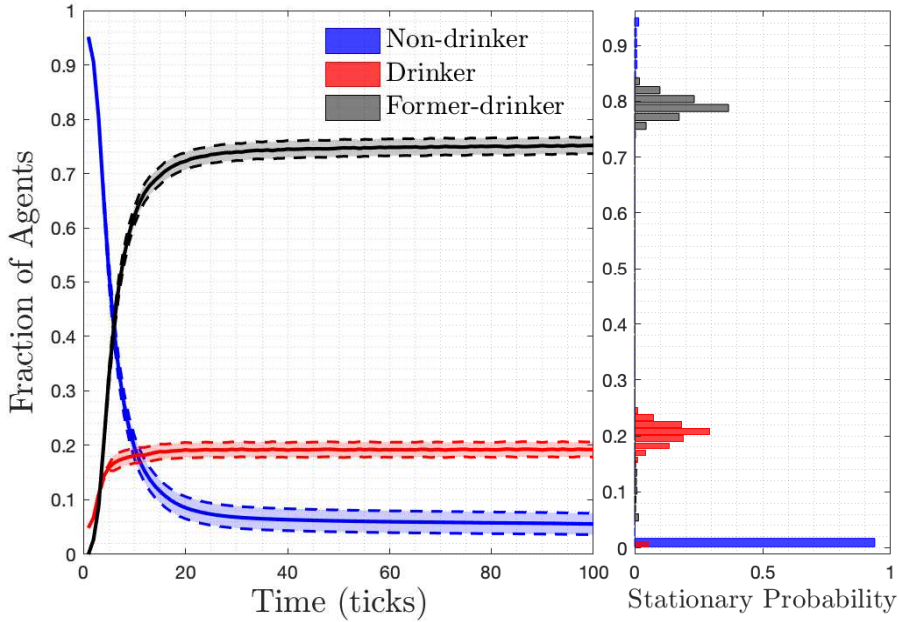}
    \caption{\textbf{Time series of fraction of agents with different status:} 
    (Left) The fractions for  average of $1000$ different stochastic simulations (solid curve) $\pm$ two standard deviations (dashed curve) of non-drinkers  (blue), drinkers (red), and former drinkers (black).  The influence rate within contexts $\beta_C$ are selected so we reach  around $20\%$ (Mean =0.1923; 95\% CI (0.1785, 0.2061) ) prevalence of drinkers at quasi-stationary state.  (Right) The stationary distributions of fractions with different status  for $1000$ simulated sample paths (histograms).}
    \label{fig:ts}
\end{figure}
In order to  estimate unknown  drinking influence success parameter, $\beta_{C_i}$, we calibrate transmission probability  to the current drinking prevalence using   Method of Simulated  Moments (MSM) \cite{mcfadden1989method}. 
We define  context related transmission probability as $\Tilde{\beta}_{C_i}=\sigma_{C_i}\beta_{C_i}$, and assume this parameter is the same for all contexts. We, therefore,  choose the unknown parameter as $\Tilde{\beta}=\sigma\beta$ and the first moment, mean of the prevalence at quasi-stationary state, as $20\%$ of drinkers,  which is estimated  based on the survey data explained in Section \ref{method}. After finding the optimized  value $\Tilde{\beta}$, the reported drinking influence successes in Table (\ref{tab:par-def-value}) are defined as $$\beta_{C_i}=\frac{\Tilde{\beta}}{\sigma_{C_i}}.$$
The Figure (\ref{fig:ts}) illustrates the typical progression of the drinking behavior to reach the current prevalence of $20\%$ drinkers among colleague students in Tempe campus, ASU. 
\subsection{Sensitivity analysis}
One set of the parameters in the model with less precise estimation is context related parameter: contact probability at context $\sigma_C$ and drinking influence success $\beta_C$. Combining these two parameters, we define transmission probability at context $C$  as $\Tilde{\beta}_{C}=\sigma_C\beta_C$ and  select it as inputs for the Sensitivity Analysis (SA), which is called Parameter of Interest (POI).
As Quantity  of Interest (QOI) we select  fraction of drinkers at quasi-stationary state.
\begin{figure}[htp]
    \centering
   \includegraphics[width=.65\textwidth]{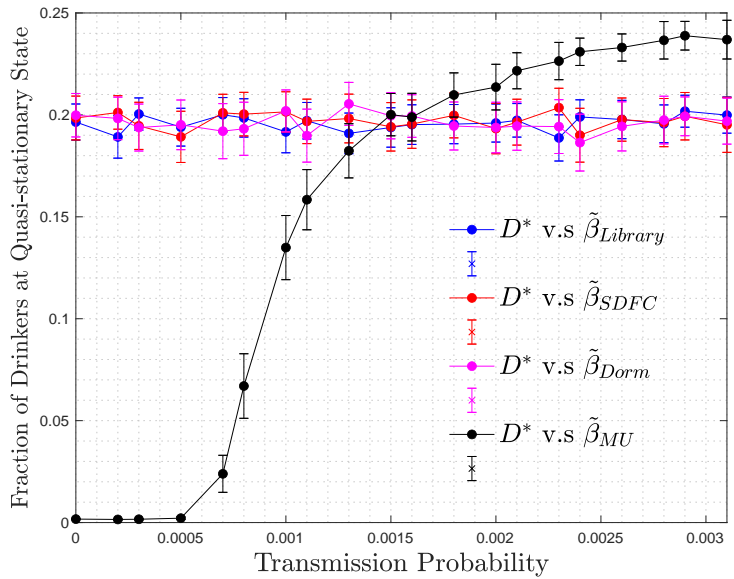}
\caption{ \textbf{Local SA on transmission probability at different contexts (the x-axis) against drinking prevalence at quasi-stationary state 
(y-axis):} The  curve  corresponding to  context $C$ is the plot of fraction of drinkers at quasi-stationary state (y-axis $=D^*$) as the  transmission probability of corresponding context (x-axis $=\tilde{\beta}_{C}$) changes while  the transmission probability for the other contexts than context $C$ are  constant at their baseline value. The drinking  prevalence are averaged over the $200$  simulations with reported $95\%$ confidence intervals.   While transmission probability in other contexts does not effect drinking prevalence, the transmission probability for  MU is highly effective. The retardation in  growth rate in MU  indicates the assumption of sigmoidal pattern: when $\tilde{\beta}_{MU}$ is big enough, that is for $\tilde{\beta}_{MU}\geq 0.001$ drinking prevalence increases rapidly and level of at its plateau of around $25\%$.}
\label{fig:lsa}
\end{figure}
 Conducting the pretest of $1000$ runs at parameter values defined  at baseline and recording output at every tick, we observed that from  $t=50$ to $t=100$
the number of drinkers  have stabilised with some random fluctuations around its  mean,  Figure (\ref{fig:ts}). Therefore, we used that $50$ ticks as the  reasonable time period for the sensitivity analysis \cite{ten2016sensitivity}.

%\subsubsection{Local SA of context influences }
  We first conducted a local SA on POIs by changing one parameter at a time to evaluate their impact on the QOI. That is, over its reasonable range we vary one POI- while keeping other parameters at their baseline value- and observe  its impact on QOI. Using NetLogo’s Behavior Space feature we programmed a series of parallel runs with a given range of parameter values, to carry out local sensitivity.
  In a given  set of simulations, the code was run for a
  transmission probability at a context $C$ in $[0,2\tilde{\beta}_C]$ per tick. The reported result in Figure (\ref{fig:lsa}) is the average of $200$ simulations.
In all contexts except  MU the QOI is  almost unchanged as   POI increases in its range. But the nonlinear effect in the experiment for MU  is evident of the logistic-shaped curve:  there is a tipping point below which increasing transmission probability does not affect drinking prevalence drastically, and above which  transmission probability would bring the drinking behavior outbreak stabilizes at $25\%$. 
The local SA for our model  indicates that $\beta_{MU}$ is by far the most effective parameter for resulting  an outbreak or bringing  the epidemic under control.  The reason is that  the density of the population in this context is higher than the other contexts, Figure (\ref{fig:visiting}).

% \subsubsection{Global sensitivity analysis of context influences}

We then used \textit{nlrx} package in R \cite{salecker2019nlrx} to conduct   two different global SA: Regression-based  method and  Variance-based Sobol methods. 

\noindent\underline{\textit{ Regression-based method}}: Assuming monotonicity between inputs and output, Partial  Correlation Coefficient (PCC) and Standard Regression Coefficients (SRC) are reliable  methods that provide a measure of this monotonicity \cite{marino2008methodology}. 
With the assumption  that there is no correlation between POIs, PCC measures the  degree of correlation between QOI and one POI by removing the effect of other POIs on QOI \cite{iman1979use}. SRCs are the  coefficient for the regression model that are  taken from standardised  POIs and QOI\cite{iman1979use}.  Therefore, higher values for PCC and SRC indicates more sensitivity to the corresponding POI.

We combined these regression methods with Latin Hypercube Sampling (LHS) \cite{marino2008methodology, blower1994sensitivity}
to conduct an SA on our model output. Via LHS sampling, we selected $1000$ samples from a uniform distribution of the POI ranges in $[0,2\tilde{\beta}_C]$ for all four contexts and run the model for each sample to calculate PCCs and SRCs for QOI, the Table (\ref{GSA}). Similar to local SA, the result indicates that transmission probability in MU is highly sensitive parameter with its index is around $0.8$ for both PCC and SRC, while the values for other contexts is almost negligible.  Producing the same rankings for PCC and SRC is the result of the assumption of no correlation  imposed on the POIs \cite{iman1985fortran}.

\noindent\underline{\textit{ Variance-based method}}:
Despite being costly in computation\footnote{The model ran for at least $4\times (1000+2)=4008$ times to generate sufficient output to calculate  variances for four parameters of transmission probabilities and $1000$ sample for each of them \cite{saltelli2008global}.}, Sobol method is a model-free variance-based method capable of quantifying the effect of POIs and their interaction on the variance of QOI with the  assumption that POIs are independent \cite{saltelli2008global}.

We carried out Sobol SA using \textit{nlrx} package for a  total sample size of $1000$ runs, sampled from uniform distributions in the range $[0,2\tilde{\beta}_C]$. We used bootstrap confidence intervals for $100$ bootstrap samples to assess the  accuracy of the estimated sensitivities.
 The resulting main effect  and total-order indices with their $95\%$ confidence intervals are shown in Table (\ref{GSA}).  
A number of confidence intervals contains negative values that are  biologically meaningless but  are the result of numerical inaccuracy of the method \cite{saltelli2008global}. We could have discarded these negative ranges by setting negative indices equal to zero but it triggers a bias in bootstrap confidence interval, so we kept negative values.  The sum of all main effect indices is $0.81$, that means more than $80\%$ of drinking prevalence variance is  explained by sole impact of transmission probabilities and less than $20\%$ explained by their interaction. The total order indices indicate that all the mentioned ineffective parameters in previous Sections   contribute to a tangible variance after including interaction effects. Additionally, the  parameters have  wide confidence intervals (order of magnitude $-1$) showing that  their estimate needs to be improved by increasing the number of bootstrap samples.
Finally, similar to previous results in SA, $\tilde{\beta}_{MU}$ is  the most sensitive parameter as its main effect and total order indices are larger than those of other parameters. 

\begin{table}[htp]
\centering
\resizebox{\columnwidth}{!}{\begin{tabular}{lcccl}\toprule
& \multicolumn{2}{c}{\textbf{Regression-based Method}} & \multicolumn{2}{c}{\textbf{Variance-based Method}}
\\\cmidrule(lr){2-3}\cmidrule(lr){4-5}
           & PCC (95\% CI)  & SRC (95\% CI)&  Main Effect (95\% CI) & Total Order (95\% CI) \\\midrule
$\tilde{\beta}_{Dorm}$    & $0.098$($[0.028,0.158]$)  & $0.042([0.016,0.069])
$& $0.043$($[0.000,0.086]$) &  $0.182([-0.018,0.210])$ \\
$\tilde{\beta}_{Library}$  & $ -0.040([-0.098,0.023]) $&  $-0.017 ([-0.039, 0.005])$  & $0.041$($[0.000,0.081]$) & $0.172 ([-0.037,0.246])$ \\
$\tilde{\beta}_{MU}$ & $0.905([0.888,0.924])$  & $0.890  [(0.866,0.913])$ &  $0.695 ([0.683,0.706])$ &  $0.832 
([0.744,0.889])$\\
 $\tilde{\beta}_{SDFC}$  & $0.076([0.002,0.129])$   &  $0.032 ([0.008,0.058])$ & $0.041([0.000,0.083])$ & $0.175([-0.022, 0.226]) $ \\\bottomrule
\end{tabular}}
\caption{\textbf{ Sensitivity ranking of Partial Correlation Coefficients (PCC) and Standard Regression Coefficients (SRC) and Main effect  and total order Sobol indices for for transmission probabilities}:  Ten replications of a LHS design with $1000$ samples was simulated for run time of $100$ ticks for each simulation. Transmission probability in MU, $\tilde{\beta}_{MU}$, is the most sensitive parameter, while transmission probability in other contexts has almost no impact on the fraction of drinkers at quasi-stationary state.
    Ten replications of a Sobol design with $1000$ samples and $100$ bootstrap samples was simulated for  run time of $100$ ticks for each simulation. Transmission probability in MU, $\tilde{\beta}_{MU}$, is the most sensitive parameter, and transmission probability in other contexts play a role on the fraction of drinkers at quasi-stationary state when they interact with other parameters. }
    \label{GSA}
\end{table}

\section{Discussion}

We used and analyzed the collected data of demographic characteristics, alcohol drinking behavior, and daily activity of  students and designed an  Agent-based model  to understand social behavioral 
mechanisms that drive drinking patterns  at  the Arizona  State University (ASU) students community. Our model tracks the movement and interaction of a population of non-drinker, drinker and former drinker agents on a 2-dimensional lattice environment including several patches that resemble  different contexts  of university campus.  Non-drinker agents became drinker on the basis of “context-influence” implemented  in terms  context related transmission probability, which is a combination of probability of contact and drinking influence success at any given context. A former drinker agent has also context-influence reasoning plus internal tendencies to become drinker again. And finally the drinker stops drinking after passing a certain drinking period.

Our data analysis on the sample of collage aged students at 
ASU confirmed the presence of significant peer influences (via drinker friends) as well as show how high socialization within various (and even non-drinking) social contexts impacts drinking dynamics in the simulated population, Figures  (\ref{fig:drinking_data} and \ref{tab:cart}). 
      Throughout our modeling study, we have focused on the  effects of environmental factors to drinking behavior  of ASU students. We \textbf{linked the social mixing behavior (in non-drinking) contexts to drinking patterns} and  \textbf{found out governing factors on college drinking dynamic}. Within this framework, the  sensitivity analysis explicitly showed the role  of time spent in a particular context on the dynamics of alcohol drinking. In particular, dynamics of college students drinking is primarily governed by high crowded contexts and the intensity of social mixing. %governs alcohol drinking pattern, 
      However, the relationship between mixing intensity in social contexts and drinking patterns, is non-trivial and our result (through use of ABM) suggests that drinking pattern follows a logistic trend.  Specifically, local sensitivity analysis showed us that there is a threshold transmission probability at which non-drinkers converted to  drinkers most efficiently if they spend more time in the most crowded context, Figure (\ref{fig:lsa}).

  The findings of our model study  needs to  be understood in terms of the several limitations. 
    First, our parametrization and calibrations of model parameters are subject to limited sample size of the data. 
    Second, ABMs in nature are challenging  when applied to understand interaction of agents ignoring highly subjective and not necessarily rational behaviors in the system of social science \cite{bonabeau2002agent}. Third, there may be high uncertainty in some of the considered parameters, which may depends on specific environment and context. Fourth, because the main goal of this study was to understand the role of  environment influences on alcohol drinking behaviors, we ignored many other factors (such as age, race, or gender) that may be linked to drinking. Future  research should carefully measure these factors and capture their impact via an ABM coupled with social network framework constructed based on  the correlations between the number of friends a person has and his/her daily activity within the campus. This social network should be characterized  by ages, ethnicity, social groups, and geographic location. 
    Although the model is still too simple to directly guide intervention efforts, the qualitative trends predicted by these simulations can be useful in designing studies to quantify the effectiveness of different intervention approaches.

\section*{Acknowledgement}
The authors acknowledge Annabel Judd for providing the raw data, and Calvin Pritchard, and Research Computing at Arizona State University  (Gil Speyer, Rebecca Belshe, Jason Yalim, and William Dizon) for providing simulation guidance, HPC, and storage  resources that have contributed to the research results reported within this paper. The authors also thank  Jan Salecker (the author and maintainer of \textit{nlrx}  package)  for his  useful help and suggestions on using \textit{nlrx}  package. 

%This work was supported by the endowment for the Evelyn and John G. Phillips Distinguished Chair in Mathematics at Tulane University and grants from the National Institutes of Health National Institute of Child Health and Human Development (R01HD086794) and Office of Adolescent Health (TP2AH000013) and the 
%National Institute of General Medical Sciences program for Models of Infectious Disease Agent Study (U01GM097658).
The content is solely the responsibility of the authors and does not necessarily represent the official views of the National Institutes of Health.

\bibliographystyle{unsrt}
\bibliography{alcohol_ref}
\section*{Appendix}
\begin{table}[htp]
\centering
\resizebox{.95\columnwidth}{!}{\begin{tabular}{llp{13cm}}
\tabularnewline
\hline 
\textbf{Random Variable} & \textbf{Value}& \textbf{Implications}\tabularnewline
\hline
$P_i(C_k)$& Empirical distribution  &Prob that individual $A_i$ moves to context $C_k$.\tabularnewline
\hline 
$\phi$& Uniform (0,1) &Resistency: susceptibility level of non- or  former drinker times infectivity level of dinker.\tabularnewline
\hline 
$\sigma$& Bernoulli($\sigma_{C_k}$) &Prob of making contact between two individual visiting context $C_k$.\tabularnewline
\hline 
$\beta$& Bernoulli ($\beta_{C_k}$) &Prob of drinking behavior transmission happens within one contact at  context $C_k$.\tabularnewline
\hline 
$\rho_i$& Bernoulli($\rho$) &Prob of alcohol  drinking reinitiation at discretion for former drinker $A_i$.\tabularnewline
\hline 
$T_i=1/\gamma_i$& Exponential($\gamma$) &Duration of drinking behavior for drinker $A_i$.\tabularnewline
\hline 
\end{tabular}}
\caption{\textbf{Random variables  used in ABM algorithm and their corresponding distributions.}}
\label{tab:stocastic}
\end{table}

 \begin{table}[htp]
 \centering
\resizebox{.83\columnwidth}{!}{
\begin{tabular}{llp{12cm}p{3cm}}
 \toprule[1.5pt]
    & \textbf{Symbol} & {\textbf{ Definition} }&\textbf{Range}\\
  \cmidrule(lr){2-4}\cmidrule(l){2-4}
%     %=====================Context===============%
&n&   Total number of socializing contexts.&-- \\ 
Context   &$\sigma_{C_i}$ &  Prob. of contact between two  individuals visited context $C_i$. & $[0,1]$\\  
Agent& $\beta_{C_i}$&Prob. of drinking behavior transmission per contact  at $C_i$.&$[0,1]$\\
Parameters  & $I_{C_i}(t)$&  The number of drinkers at context $C_i$ at time $t$.  & $\{0,1,\cdot\cdot\cdot,N\}$ \\
 & $\bar{N}_{C_i}$&  Per area  average number  of individuals visiting context $C_i$. & -- \\
  & $p_{C_i}(t)$&  Mean proba. of visiting context $C_i$.  &$[0,1]$ \\
   %=====================people===============%
  \cmidrule(lr){2-4}\cmidrule(l){2-4}
    & $N$&Total number of individuals. &--\\  
Individual & $R_i(t)$&  class-year of individual $A_i$ at time $t$. & $\{0 \textnormal{(Freshman)}, 1 \textnormal{(Sophomore)}$ \\
Agent &  & &$ 2 \textnormal{(Junior)},  3 \textnormal{(Senior)},  4 \textnormal{(Graduate)}\}$ \\
Parameters & $P_i$& Per tick prob. vector  of visiting contexts by individual $A_i$. $P_i$ is a vector of size n where $k^{th}$ element ($P_i(C_k)$
) is the chance that individual $A_i$ visit context $C_k$. & $[0,1]^n$ \\
& $V_i(t)$& The visited context  of individual $A_i$ at time $t$: $A_i$ is at context $V_i(t)$ at time $t$. &$\{C_1,\cdot\cdot\cdot, C_n\}$ \\
& $S_i(t)$& Drinking state of individual $A_i$ at time $t$.  & $\{ND,D,SD\}$\\
 %=====================Alcohol===============%
 \cmidrule(lr){2-4}\cmidrule(l){2-4}
Alcohol     & $\gamma$& Prob. of recovery from drinking. &$[0,1]$\\  
Drinking  & $\rho$& Prob. of re-initiation drinking at discretion.   & $[0,1]$ \\
Parameters & $\phi$& Resistancy: susceptibility of non- or former drinkers times infectivity of drinker. & $[0,1]$\\
\bottomrule[1.5pt]
 \end{tabular}}
\caption{\textbf{The set of all parameters and attributes defined in our ABM.}}
\label{tab:abm_par}
\end{table}

\begin{table}[hpt]
\centering
\resizebox{.9\columnwidth}{!}{\begin{tabular}{lp{9.2cm}}
\toprule[1.5pt]
   \textbf{Notation} & \textbf{Description}\\
  \cmidrule(lr){1-2}
  $\textbf{A}_k$ &  An individual with ID k\\
    $\eta_k(t)=(R_k(t),P_k, V_k(t), S_k(t))$ &  Attribute of  individual  $\textbf{A}_k$ at time t\\
    $S.add(m)$ & Add element m to a set S\\ 
  $DP(k)$& Drinking  period for drinker individual $\textbf{A}_k$\\
  $ern(a)$ & Exponential random number with parameter $a>0$ \\
    $urn$ & Uniform random number in $[0,1]$\\
    $rem(a,b)$&  Remainder after division of a by b\\
\bottomrule[1.5pt]
\end{tabular}}
\caption{Table of notation for a conventional model  in algorithm.}
\label{notation}
\end{table}
\begin{minipage}{.9\linewidth}
\begin{algorithm}[H]
 \SetAlgoLined
 $NC=\emptyset$ \%\text{$NC$ is an empty set.}\;
\For{$\textbf{A}_k$ with $\eta_k(t)=(R_k(t),P_k, V_k(t), S_k(t))$}
{{$\%==============${\textit{/*Movement*/}}$==============\%$\;
$j=min\{m: urn\leq \sum_{i=1}^m P_k(i)\}$\;
$V_k(t+\Delta t)\leftarrow C_j$
}\;
\If{$S_k(t)=D$}
{$\%========${\textit{/*Drinking behavior transmission*/}}$========\%$\;
\For{$\textbf{A}_j\in \{\textbf{A}_i\notin NC: V_i(t)=V_k(t), S_k(t)\neq D\}$}  
            { \If{urn$\leq \sigma_{V_k(t)}\beta_{V_k(t)}urn$}
       {{$S_j(t+\Delta t)\leftarrow D$,~~$DP(j)\leftarrow -1$ , ~~$NC.add(\textbf{A}_j)$\;}
      }
    }%end of for infected list
$\%========${\textit{/*Recovery from drinking*/}}$========\%$\;    
\uIf{$DP(k)=-1$}
           {$DP(k)\leftarrow max\{\Delta t,ern(\gamma$)\}\; }
              \uElseIf{$DP(k)>0$}
              {$DP(k)\leftarrow DP(k)-\Delta t$\;}
            \Else{$S_k(t+\Delta t)\leftarrow FD$\;}
            }
            $\%========${\textit{/*Reinitiation at discretion*/}}$========\%$\;  
            \If{$S_k(t)=FD$~\&  $k\notin NC$~\& $urn \leq \rho$}
            {$S_k(t+\Delta t)\leftarrow D$,~~
            $DP(k)\leftarrow -1$,~~$NC.add(\textbf{A}_k)$}
            $\%========${\textit{/*class-year*/}}$========\%$\;  
            \If{$rem(t,1440)=0$}
            {$R_k(t+\Delta t)\leftarrow R_k(t)+1$\;
            \If{$R_k(t+\Delta t)=5$}{$R_k(t+\Delta t)=0$,~~$S_k(t+\Delta t)=ND$}}
           }
 \caption{ The model dynamic at typical time $t$}\label{alg}
\end{algorithm}
\end{minipage}
\iffalse
\begin{figure}[htp]
    \centering
    \includegraphics[width=.48\textwidth]{LHS.eps}
    \includegraphics[width=.45\textwidth]{Sobol.eps}
    \caption{\textbf{Left: Sensitivity ranking of Partial Correlation Coefficients (PCC) and Standard Regression Coefficients (SRC) for the fraction of drinkers}:  Ten replications of a LHS design with $1000$ samples was simulated for run time of $100$ ticks for each simulation. Transmission probability in MU $\tilde{\beta}_{MU}$ is the most sensitive parameter, while transmission probability in other contexts has almost no impact on the fraction of drinkers at quasi-stationary state.
    \textbf{Right: Main effect  and total order Sobol indices for transmission probabilities:}
    Ten replications of a Sobol design with $1000$ samples and $100$ bootstrap samples was simulated for  run time of $100$ ticks for each simulation. Transmission probability in MU $\tilde{\beta}_{MU}$ is the most sensitive parameter, and transmission probability in other contexts play a role on the fraction of drinkers at quasi-stationary state when they interact with other parameters. }
    \label{fig:GSA}
\end{figure}
\fi

 \end{document}